\documentclass[conference,letterpaper]{IEEEtran}
\IEEEoverridecommandlockouts

\usepackage[ruled,vlined,linesnumbered]{algorithm2e}
\usepackage{algorithmic}
\usepackage{amsmath,amssymb,amsfonts}
\usepackage{booktabs}
\usepackage{cite}
\usepackage{graphicx}
\usepackage{hyperref}
\hypersetup{
colorlinks = true, 
linkcolor=black, 
citecolor=black, 
urlcolor=black 
}
\usepackage{multicol}
\usepackage{multirow}
\usepackage{nicefrac}
\usepackage{paralist}
\usepackage{stfloats}
\usepackage{subcaption}
\usepackage[inkscapelatex=false]{svg}
\usepackage{textcomp}
\usepackage{wrapfig}
\usepackage{xcolor}
\usepackage{xspace}

\def\BibTeX{{\rm B\kern-.05em{\sc i\kern-.025em b}\kern-.08em
T\kern-.1667em\lower.7ex\hbox{E}\kern-.125emX}}

\newcommand{\randomApproach}{\texttt{RS}\xspace}
\newcommand{\hornBro}{\texttt{HornBro}\xspace}
\newcommand{\approach}{\texttt{QRep}\xspace}
\newcommand{\unitar}{\texttt{UnitAR}\xspace}
\newcommand{\testSuite}{\ensuremath{\mathit{TS}}\xspace}
\newcommand{\testCase}{\ensuremath{\mathit{tc}}\xspace}
\newcommand{\timeBudget}{\ensuremath{\mathit{B}}\xspace}
\newcommand{\iterationBudget}{\ensuremath{\mathit{\timeBudget_i}}\xspace}
\newcommand{\remainingBudget}{\ensuremath{\mathit{\timeBudget_r}}\xspace}
\newcommand{\numFailedTestCases}{\ensuremath{\#\testCase_{\text{fail}}}\xspace}
\newcommand{\numIterations}{\ensuremath{\mathit{I}}\xspace}
\newcommand{\currentIteration}{\ensuremath{\mathit{i}}\xspace}
\newcommand{\fitnessScore}{\ensuremath{\mathit{fit}}\xspace}
\newcommand{\susScore}{\ensuremath{\mathit{sus}}\xspace}
\newcommand{\hellingerDistance}{\ensuremath{\mathit{H}}\xspace}
\newcommand{\circuit}{\ensuremath{\mathit{C}}\xspace}
\newcommand{\circuitInit}{\ensuremath{\mathit{C_{init}}}\xspace}
\newcommand{\CandPatches}{\ensuremath{\mathtt{CandPatches}}\xspace}
\newcommand{\modCircuit}{\ensuremath{\tilde{C}}\xspace}
\newcommand{\gate}{\ensuremath{\mathit{g}}\xspace}
\newcommand{\patch}{\ensuremath{\mathit{p}}\xspace}
\newcommand{\RepairedCircuits}{\ensuremath{\mathit{RepairedCircuits}}\xspace}
\newcommand{\TotalFaultyCircuits}{\ensuremath{\mathit{TotalFaultyCircuits}}\xspace}
\newcommand{\numQubits}{\ensuremath{\#\mathit{qubits}}\xspace}

\begin{document}
\title{Quantum Circuit Repair by Gate Prioritisation\thanks{E. Mendiluze Usandizaga is supported by Simula's internal strategic project on quantum software engineering. T. Laurent is supported in part with the financial support of grant 13/RC/2094\_2 to Lero - the Research Ireland Research Centre for Software. P. Arcaini is supported by the ASPIRE grant No. JPMJAP2301, JST. S. Ali is supported by the Qu-Test project (Project \#299827) funded by the Research Council of Norway and Oslo Metropolitan University's Quantum Hub.}}


\author{\IEEEauthorblockN{Eñaut Mendiluze Usandizaga\IEEEauthorrefmark{1}\IEEEauthorrefmark{2},
Thomas Laurent\IEEEauthorrefmark{3},
Paolo Arcaini\IEEEauthorrefmark{4}, and
Shaukat Ali\IEEEauthorrefmark{1}\IEEEauthorrefmark{2}}
\IEEEauthorblockA{\IEEEauthorrefmark{1}Simula Research Laboratory, Oslo, Norway}
\IEEEauthorblockA{\IEEEauthorrefmark{2}Oslo Metropolitan University, Oslo, Norway}
\IEEEauthorblockA{\IEEEauthorrefmark{3}RI Lero \& Trinity College Dublin, School of Computer Science and Statistics, Dublin, Ireland}
\IEEEauthorblockA{\IEEEauthorrefmark{4}National Institute of Informatics, Tokyo, Japan}
}

\maketitle

\begin{abstract}
Repairing faulty quantum circuits is challenging and requires automated solutions. We present \approach, an automated repair approach that iteratively identifies and repairs faults in a circuit. \approach uniformly applies patches across the circuit and assigns each gate a suspiciousness score, reflecting its likelihood of being faulty. It then narrows the search space by prioritising the most suspicious gates in subsequent iterations, increasing the repair efficiency. We evaluated \approach on 40 (real and synthetic) faulty circuits. \approach completely repaired 70\% of them, and for the remaining circuits, the actual faulty gate was ranked within the top 44\% most suspicious gates, demonstrating the effectiveness of \approach in fault localisation. Compared with two baseline approaches, \approach scales to larger and more complex circuits, up to 13 qubits.
\end{abstract}


\section{Introduction}

Quantum software testing has become an active area of research with several techniques proposed to detect faults~\cite{qseRoadmapTOSEM2025,quantumTestingRoadmapTOSEM2025,CTQuantumQRS2021,LongJSS2024,genTestsQPSSBSE2021,honarvar2020property,quraTestASE23,metamorphic,muqeet2024mitigating,mutation-based,pauliStringsASE2024,projectionBased,quitoASE21tool,testingQuantumICST2021}. However, little attention has been paid to quantum circuit repair, despite the need for automated and scalable approaches to fix faults in increasingly complex quantum circuits. Some approaches have been proposed for quantum circuit repair. \hornBro~\cite{tan2025hornbro} simplifies circuits to Clifford gates and frames repair as parameter tuning, while \unitar~\cite{unitar} generates unitary operator patches from a black-box algebraic model. Both face scalability challenges: \hornBro adds many new gates, causing scalability issues and potentially failing to fully convert the circuit in Clifford gates, and \unitar cannot handle more than five qubits. Guo et al.~\cite{guo2024repairing} explore semi-automatic repair with ChatGPT, but it relies on human input and struggles to generalise beyond its training circuits. 



To address key limitations in the current literature regarding practicality and usability, we propose \approach, a quantum circuit repair approach that combines fault localisation with quantum circuit repair via gate-based modifications. As shown by Ishimoto et al.~\cite{mutationBasedFaultLoc}, quantum gate-based modifications are more effective at identifying quantum faults than classical mutations. Building on this insight, \approach automatically repairs circuits and, when it does not fully repair the circuit, it assists users in localising the faults and generating patches that can guide manual repair.

\approach has two phases. First, it localises faulty gates by assigning a {\it suspiciousness score} to each gate. Second, this score guides the prioritisation of repair patches, focusing on the most suspicious gates. This strategy reduces the search space and avoids unnecessary patch attempts. We performed a preliminary empirical evaluation with 40 faulty quantum circuits. Results show that \approach can repair circuits of up to 13 qubits, compared to existing methods that typically handle up to 4 qubits. Even when a circuit is not fully repaired, \approach improves it, and ranks the most suspicious gates, reducing the search space for manual repairs.

\section{Background}

Quantum software presents unique challenges as it utilises phenomena from quantum mechanics, such as {\it superposition} and {\it entanglement}, to perform computations~\cite{nielsen2010quantum}. Superposition allows a {\it qubit} (or {\it quantum bit}) to be in multiple states simultaneously, meaning that, in contrast to a classical bit, it can exist in both 0 and 1 states at the same time. {\it Entanglement} is the phenomenon where two or more qubits are correlated, regardless of their distance. These and other quantum phenomena enable quantum computers to process large amounts of data very efficiently, theoretically allowing them to solve problems that classical computers cannot handle.

\begin{wrapfigure}[12]{r}{0.3\linewidth}
\centering
\includegraphics[width=\linewidth]{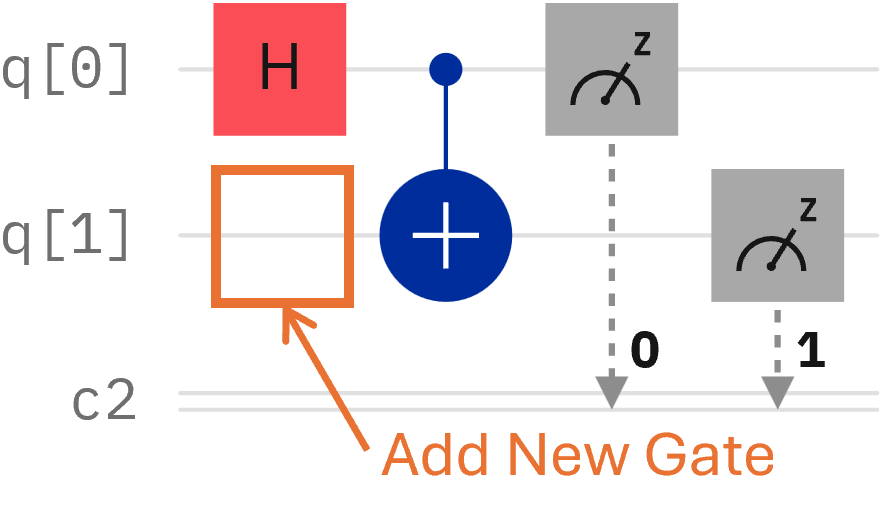}
\caption{Quantum circuit and patch example. Qubits: $q[0]$ and $q[1]$; group $c_{2}$ of two classical bits.}
\label{fig:example}
\end{wrapfigure}
Quantum software is typically implemented as a {\it quantum circuit}, which consists of a set of qubits and {\it quantum gates} that are applied to alter their states. Quantum gates serve as the quantum equivalent of logic gates in the classical domain. These gates can change the state of single qubits or multiple qubits simultaneously. Each gate in the circuit affects a qubit's state in a specific way. Fig.~\ref{fig:example} shows a two-qubit entanglement circuit, with a Hadamard gate $H$ applied to the first qubit, creating a superposition state, and a CNOT multi-qubit conditional gate, illustrated as a dot and a cross. At the end of the circuit, two measurement operations translate the quantum state into a classical state.

\section{\approach{} -- Proposed Repair Approach}


Alg.~\ref{alg:overview} shows an overview of \approach. Its inputs are:
\begin{inparaenum}[(i)]
\item a faulty quantum circuit \circuitInit to be repaired,
\item a failing test suite \testSuite, i.e, a test suite with at least one failing test case \testCase,
\item a time budget \timeBudget for the entire repair process, and
\item the number of iterations \numIterations the repair process must go through.
\end{inparaenum}
\begin{algorithm}[tb]
\scriptsize
\caption{\approach Automated Repair Overview}
\label{alg:overview}
\KwIn{Faulty quantum circuit \circuitInit; failing test suite \testSuite; total time budget \timeBudget; number of iterations \numIterations}
\KwOut{Best patches, gate suspiciousness ranking}
\vspace{0.1cm}
\tcp{Initialization}
Evaluate \circuitInit on \testSuite and compute initial fitness as in Eq.~\ref{eq:fitness}\;
\label{algline:initialization}
\vspace{0.1cm}
\tcp{Fault Localisation (see Sect.~\ref{sec:faultLocalisation})}
\For{gate \gate in \circuitInit}{\label{algline:forGate}
Remove gate \gate from \circuitInit obtaining \modCircuit\;\label{algline:removegate}
Evaluate candidate circuit \modCircuit over \testSuite\;\label{algline:evaluateCirc1}
\If{\testSuite passes over \modCircuit}{\label{algline:ifTSpass1}
\Return \modCircuit as completely repaired circuit\;\label{algline:returnFix1}
}
Compute fitness $\fitnessScore(\modCircuit)$ as in Eq.~\ref{eq:fitness}\;\label{algline:fitCandCirc1}
Update gate suspiciousness $\susScore(\gate)$ as in Eq.~\ref{eq:suspiciousness}\;\label{algline:updateSus1}
}
\vspace{0.1cm}
\tcp{Patch Generation (see Sect.~\ref{sec:patchGen})}
Generate candidate patches \CandPatches\;\label{algline:generatePatches}
Distribute patches \CandPatches\;\label{algline:distributePatches}
\vspace{0.1cm}
\tcp{Repair (see Sect.~\ref{sec:repair})}
Calculate remaining time: \remainingBudget $=$ \timeBudget $-$ fault localisation time\;\label{algline:remainingTime}
Calculate iteration time budget \(\iterationBudget = \remainingBudget/\numIterations\)\;\label{algline:setItTime}
\For{\currentIteration in \numIterations}{\label{algline:whilePatch}
init $\mathit{timer}$\;
\While{$\mathit{timer} < \iterationBudget$ AND patches exist in \CandPatches}{\label{algline:whileIt}
Apply patch \patch from \CandPatches to \circuitInit obtaining \modCircuit\;\label{algline:applyPatch}
Evaluate candidate circuit \modCircuit\;\label{algline:evaluateCirc2}
\If{\testSuite passes over \modCircuit}{\label{algline:ifTSpass2}
\Return repaired quantum circuit \modCircuit\;\label{algline:returnFix2}
}
Compute fitness $\fitnessScore(\modCircuit)$ as in Eq.~\ref{eq:fitness}\;\label{algline:fitCandCirc2}
Update gate suspiciousness $\susScore(\gate)$ as in Eq.~\ref{eq:suspiciousness}\;\label{algline:updateSus2}
}
Keep in \CandPatches the most suspicious gates (top $\left(1 - \frac{\currentIteration}{ \numIterations}\right)\%$) \;\label{algline:selectGates}
}
\Return best patches, gate suspiciousness ranking\;\label{algline:returnNotFix}
\end{algorithm}

\approach uses an iterative process that prioritises the most suspicious gates in each iteration. Between iterations, the least suspicious gates are discarded,  progressively reducing the search space, while prioritising the most suspicious gates.

In Line~\ref{algline:initialization}, \approach executes the test suite \testSuite on the faulty circuit \circuitInit. Then, it computes an initial fitness score $\fitnessScore(\circuitInit)$, calculated based on the number of failed test cases \numFailedTestCases, and the total Hellinger distance \hellingerDistance of all test cases from the ideal distribution. Formally, the fitness function that must be minimised is defined as:
\begin{equation}
\fitnessScore(\circuit) = \numFailedTestCases + \sum_{i=1}^{|\testSuite|} \hellingerDistance(\testCase_i)
\label{eq:fitness}
\end{equation}

The Hellinger distance quantifies the similarity between two probability distributions, from 0 (identical) to 1 (totally different)~\cite{li2024uncertainty}. In our context, it quantifies how far the measured quantum state is from the expected state, helping identify faults and guide fitness improvement while prioritising test case reduction. 

\subsection{Fault Localisation}\label{sec:faultLocalisation}
In this step, \approach assesses the {\it suspiciousness} of the gates by removing one gate at a time from \circuitInit and evaluating the effect of its removal (Lines~\ref{algline:removegate}-\ref{algline:evaluateCirc1}).
If a modified circuit \modCircuit passes the test suite \testSuite, \approach exits and reports \modCircuit as the completely repaired circuit (Lines~\ref{algline:ifTSpass1}-\ref{algline:returnFix1}). Instead, if any test case fails, the process continues until all gate removals have been evaluated.

After each evaluation, the fitness score of \modCircuit is computed (Line~\ref{algline:fitCandCirc1}) and the suspiciousness score of the gate \gate that has been removed in \modCircuit is updated (Line~\ref{algline:updateSus1}); the score considers the change in fitness caused by the circuit modification:
\begin{equation}
\susScore(\gate) = \susScore(\gate) + \big(\fitnessScore(\circuit) - \fitnessScore(\modCircuit)\big)
\label{eq:suspiciousness}
\end{equation}

All gates are initialised with $\susScore(\gate) = 0$. The score may become \emph{negative} if modifying a gate increases the fitness.

This process assumes that if a modified gate is neither the faulty one nor related to the fault, then introducing a new modification will act as a second fault, leading to more failed test cases or greater test case failure distance, thus increasing the fitness. Instead, we assume that removing the faulty gate will reduce the effect of the fault, thereby decreasing the fitness. Over multiple modifications, gates that consistently decrease the fitness value accumulate higher suspiciousness scores, whereas gates that increase the fitness score may acquire negative scores, making them less suspicious.

\subsection{Patch Generation}\label{sec:patchGen}

In Line~\ref{algline:generatePatches}, \approach starts the patch generation by generating the list of all possible patches \CandPatches available for the faulty circuit \circuit.
This patch list contains gate-based modifications to apply to the circuit. As previously proposed in the literature~\cite{mutationBasedFaultLoc}, we can either {\it add} a new gate to the quantum circuit or {\it replace} a gate with another gate.
Every possible modification will be treated as a new candidate patch. Fig.~\ref{fig:example} shows an example of adding a new gate to the circuit.

The generated patches are then distributed uniformly across the circuit to ensure early coverage of the circuit's gates (Line~\ref{algline:distributePatches}). The uniform distribution is achieved by selecting one patch at each position in the circuit, alternating between adding and replacing gates. Also, the approach ensures that a variety of Qiskit-supported gates are used, covering different quantum gates throughout the process. This approach ensures that different gates are targeted in each iteration, enabling broad exploration of potential patches while maintaining balanced coverage across the entire circuit.

\subsection{Repair}\label{sec:repair}
Once all possible patches are generated, the iterative process begins. In Line~\ref{algline:setItTime}, the approach sets a time budget \iterationBudget for each iteration, which is determined by distributing the remaining time \remainingBudget for the overall approach (computed at Line~\ref{algline:remainingTime}) across the desired number of iterations  \numIterations.

In each iteration's inner loop (Line~\ref{algline:whileIt}), for each patch \patch, the modification is applied to the faulty circuit \circuitInit and removed from the patch list \CandPatches, generating a candidate circuit \modCircuit incorporating the modification (Line~\ref{algline:applyPatch}) that is further evaluated with the test suite \testSuite (Line~\ref{algline:evaluateCirc2}).
If a parametrised gate is introduced, the COBYLA optimiser (widely used in quantum computing) is employed to find the best parameter values. Any patch passing \testSuite is considered successful, marking the circuit as repaired, and terminating the process (Lines~\ref{algline:ifTSpass2}-\ref{algline:returnFix2}).

After the execution of \testSuite, the fitness of \modCircuit is calculated using Eq.~\ref{eq:fitness} (Line~\ref{algline:fitCandCirc2}). Then, in Line~\ref{algline:updateSus2}, the suspiciousness score of the related gate \gate is updated using Eq.~\ref{eq:suspiciousness}.

If the iteration concludes without finding a successful patch and not all patches have been executed, the process proceeds to the next iteration. During this transition, the gate suspiciousness list and the current iteration \currentIteration will be used to select the most suspicious gates (Line~\ref{algline:selectGates}), and only patches linked to these gates will proceed to the next iteration. 

If, after an iteration, no more patches are available in \CandPatches or the budget has expired, the circuit will be considered as ``Not Fixed'', and \approach will report the best patches identified throughout the evaluation, along with the gate suspiciousness ranking (Line~\ref{algline:returnNotFix}). These patches are the most effective modifications to the circuit for addressing the faults and can facilitate future manual repair.

\section{Experiment Design}
We here describe the experiment design. The implementation and all results are available in our repository~\cite{ZenodoRepo}.

\subsubsection*{\bf Research questions}
To assess the effectiveness of \approach, we address the following research questions:
\begin{compactitem}
\item[\textbf{RQ1}:] \textit{How effective is \approach in repairing quantum circuits?}
\item[\textbf{RQ2}:] \textit{In the non-repaired quantum circuits, what is the partial repair effectiveness and how accurate is the fault localisation?} 
\end{compactitem}

\subsubsection*{\bf Subject Systems} 
We used 4 real faulty benchmarks from Bugs4Q~\cite{zhao2021bugs4q} and used to evaluate \unitar by Li et al.~\cite{unitar}, with qubits ranging from 2 to 4, depths from 4 to 11, and gates from 3 to 14. These faulty benchmarks originate from small quantum circuits in terms of qubit size and complexity, which is usually measured by the circuit's depth and the number of gates. Therefore, we chose to utilise two quantum mutation tools currently available in the literature, Muskit~\cite{Mendiluze2021} and QMutPy~\cite{QmutPy}, to create additional faulty circuits. We selected a set of 12 quantum circuits\footnote{Amplitude Estimation (ae); Deutsch-Jozsa (dj); Greenberger-Horne-Zeilinger State (ghz); Graph State (graphstate); Grover Search(grover); Quantum Approximate Optimization Algorithm (qaoa); Quantum Fourier Transform (qft); Quantum Phase Estimation Exact (qpeexact); Quantum Walk (qwalk); Variational Quantum Eigensolver (vqe); W-State (wstate).}, each with a different number of qubits (2-13), gates (14-50) and depth (7-38). These circuits were sourced from MQT Bench~\cite{mqtbench}, a widely recognised benchmark suite for quantum circuits. The selection of these circuits was guided by the grouping methodology outlined in~\cite{MendiluzeUsandizaga2025}, which categorises quantum algorithms based on their building blocks. We generated all possible mutants using both tools and merged the mutant sets by comparing them to avoid duplication. The mutants were then grouped by mutation operator. For each algorithm, we randomly selected a mutant from each group, resulting in three mutants per algorithm. This process resulted in a benchmark set of 36 faulty circuits.

\subsubsection*{\bf Baseline approaches}
We compare \approach with two baseline repair approaches to provide a fair assessment of its effectiveness. First, we compare it to a random search strategy (\randomApproach): this strategy randomly generates patches without guidance on suspiciousness or any patch selection strategy. Since \approach includes an optimisation algorithm for parametric gates, we allocate the same number of evaluations used for the optimiser to \randomApproach as the budget for assessing them.

Next, we compared \approach with a well-established approach from the literature called \unitar~\cite{unitar}. This approach offers strong theoretical foundations for patch generation based on algebraic models of unitarity operations. In this comparison, we adhered to the parameters and budget specified in \unitar's documentation and translated those to our approach. 

Each approach was executed with a budget of 2 hours per faulty quantum circuit, and for \unitar, a unitary matrix budget of 100,000, as in the original study.

\subsubsection*{\bf Test Suite}
For each faulty circuit \circuitInit, we construct a test suite \testSuite based on the input coverage proposed by Wang et al.~\cite{quitoASE21tool}, where we generate all possible classical input initialisations of the qubits. Since a quantum circuit can be measured in multiple bases, we create test cases for each major measurement bases, $X$, $Y$ and $Z$. Therefore, the total number of tests in \testSuite is \(2^{\numQubits} \times 3\), where \numQubits is the number of qubits of \circuitInit. Note that, for execution in \unitar, the provided implementation did not support multiple base measurements, so we ran \unitar using the default $Z$ base measurements. The differences in the test suite may lead \unitar to consider some of the generated faults as equivalent, since the faults might not be apparent in only the $Z$-basis. We chose to proceed by running \approach alongside \randomApproach across the three bases, as we believe this offers a more comprehensive view of the faults and actual repairs. In the case of \unitar, the faults may not have been completely repaired; they simply may not be visible in the $Z$-basis.\footnote{Note that this favours \unitar in the comparison with \approach. Therefore, it does not affect the soundness of our experiments.}

To determine whether a test case passes or fails, we used an oracle-based assessment method~\cite{MendiluzeUsandizaga2025}. A test case is considered failed in either of the following cases:
\begin{inparaenum}[(i)]
\item the observed output is not among the expected ones, indicating incorrect program behaviour, or
\item the output probabilities deviate significantly from the expected probabilities. 
\end{inparaenum}

\subsubsection*{\bf Experimental Setup} \label{subsec:experimentalsetupexecution}
The experimental environment consisted of a Windows machine with 20 CPUs and 64 GB of RAM.
All circuits are written in QASM 2, and we used Qiskit ver. 1.1.0 to load and modify the circuit. To execute the quantum circuits, we used Qulacs ver. 0.6.10 as the simulator. To enhance consistency and reproducibility, a fixed random seed was used during the simulations. To address the uncertainty in quantum computing, we performed a number of shots proportional to the number of qubits, calculated as $2^{\numQubits} \times 2$, that should allow us to cover all output states if necessary. \looseness=-1 

\section{Results and Analysis} \label{sec:results}
\subsubsection*{\bf RQ1 (Repair Effectiveness)}

Tab.~\ref{tab:resultsRQ1} presents the results. Each column shows the repair rate for each approach, indicating the number of circuits completely repaired out of the total number of faulty quantum circuits.
\begin{table}[tb]
\centering
\caption{RQ1 -- Number of faulty quantum circuits repaired by approach ($\RepairedCircuits/\TotalFaultyCircuits$).}
\label{tab:resultsRQ1}
\begin{tabular}{lrccc}
\toprule
\multirow{2}{*}{\textbf{Quantum Circuit}} & \multirow{2}{*}{\textbf{\# Qubits}} & \multicolumn{3}{c}{\textbf{Approach}}\\\cmidrule{3-5}
& & \textbf{\unitar} & \textbf{\randomApproach} & \textbf{\approach}\\
\midrule
Bugs4Q-id25 & 3 & 1/1 & 0/1 & 0/1\\
Bugs4Q-id26 & 2 & 1/1 & 1/1 & 1/1\\
Bugs4Q-id27 & 3 & 0/1 & 1/1 & 1/1\\
Bugs4Q-id39 & 4 & 0/1 & 0/1 & 0/1\\
QNN & 2 & 0/3 & 2/3 & 3/3\\
AE & 3 & 0/3* & 1/3 & 3/3\\
Grover & 4 & 0/3* & 1/3 & 3/3\\
Qwalk & 5 & 0/3* & 0/3 & 3/3\\
QAOA & 6 & 0/3* & 0/3 & 1/3\\
VQE & 7 & 0/3* & 0/3 & 3/3\\
QPE & 8 & 0/3* & 0/3 & 2/3\\
QFT & 9 & 0/3* & 0/3 & 2/3\\
W-state & 10 & 0/3* & 0/3 & 2/3\\
DJ & 11 & 0/3* & 0/3 & 2/3\\
Graphstate & 12 & 0/3$^\dagger$ & 0/3 & 1/3\\
GHZ & 13 & 0/3* & 0/3 & 1/3\\
\midrule
\textbf{Total} & - & \textbf{2/40} & \textbf{6/40} & \textbf{28/40} \\
\bottomrule
\end{tabular}
\\\smallskip
\scriptsize*No unitary matrixes found, $\dagger$Equivalent program (Only $Z$ basis measurements)
\end{table}

All three approaches achieve the same effectiveness in repairing real faults from Bugs4Q, each repairing 2 out of 4. However, there is a slight difference in terms of which faults are repaired. All three approaches successfully repair the id26 faulty circuit, but \unitar also repairs id25, while \approach and \randomApproach repair id27. None of the approaches can repair the id39 fault, highlighting a possible scalability issue in quantum circuit repair, as it involves a higher qubit count.

When considering the generated faulty circuits, we note that scalability is a key challenge. The artificially generated faults originate from larger, more complex circuits, resulting in a significant decrease in complete repairs for both baselines. In contrast, \approach successfully repairs 26 out of the 36 artificially generated faults, bringing the total number of repaired faults to 28 out of 40, representing a 70\% complete repair rate.

Results show the weakness of \unitar, as it is unable to repair any of the generated faults within the provided budget.
For the largest circuits, this limitation prevents even selecting unitary matrix operations, as scaling becomes increasingly challenging with the number of qubits and gates. 

\subsubsection*{\bf RQ2 (Partial Repair \& Fault Localization)}
For the non-repaired circuits, \approach provides a list of best patches and a rank of gates by suspiciousness. Thus, we analyse the partial repair of the best patch by measuring the improvement in fitness, and we evaluate how well \approach localises faults by reporting the faulty gate's rank in the gate suspiciousness list. Tab.~\ref{tab:resultsRQ2} shows the results obtained for the circuits that were not completely repaired. The second column shows the percentage improvement in circuit fitness. The last column indicates the final rank of the faulty gate within the suspiciousness list: 0\% corresponds to the top of the list (most suspicious) and 100\% to the bottom (least suspicious).
\begin{table}[tb]
\centering
\caption{RQ2 -- Fitness improvement and faulty gate location in avg across non-repaired faults from the same circuit.}
\label{tab:resultsRQ2}
{
\setlength{\tabcolsep}{5.5pt}
\begin{tabular}{lrrr}
\toprule
\textbf{Quantum Circuit} & \textbf{\#Faults} & \textbf{Improvement \%} & \textbf{Location (Top \%)}\\
\midrule
Bugs4Q-id25 & 1 & 3.6\% & 50\% \\
Bugs4Q-id39 & 1 & 10.6\% & 0\% \\
QAOA & 2 & 99.3\% & 40.5\% \\
QPE & 1 & 33.7\% & 58.2\% \\
QFT & 1 & 83.6\% & 0\% \\
W-state & 1 & 52.8\% & 25\% \\
DJ & 1 & 0.0\% & 37.2\% \\
Graphstate & 2 & 15.8\% & 29.7\% \\
GHZ & 2 & 40.2\% & 62.9\%\\
\bottomrule
\end{tabular}
}
\end{table}

We observe that although the faults were not fully repaired, all cases except one show at least some improvement in fitness. Some cases showed a significant repair rate, up to 99\%. In the case where the fitness was not reduced, the faulty gate was ranked in the top 37\% of the most suspicious gates. This means that if a user checks the gates in order of their suspiciousness, they could significantly reduce the search space or the time required to locate the fault by more than half. Some cases located the gate at the top of the suspiciousness list 0\%, meaning \approach was able to identify the faulty gate as the most suspicious. Even in the worst result, the faulty gate is still 62\% from the top, indicating that it can also help reduce the search space to some extent. Together with the provided partial repair rates, this significantly reduces the manual work required to repair the quantum circuit.

\section{Discussion and future work}
\approach shows promise for repairing and localising faults in quantum circuits. However, it supports only one modification at a time. Thus, further work is required to investigate \approach's effectiveness for circuits with multiple faults and potential improvements with multiple modifications. 

In our case, the correctness of the repaired circuit is evaluated using the provided test suite, as \approach considers a fault repaired once the circuit passes all test cases in the test suite. However, there may be test cases outside the test suite in which the repaired circuit fails. 

Regarding empirical evaluation, we evaluated the \approach on 40 faulty quantum circuits, 36 of which were artificially generated. In the future, we aim to evaluate our approach with a broader benchmark of real faults.


\section{Conclusions} \label{sec:conclusionAndRelated}
We introduced \approach, a quantum fault localisation and repair method. Preliminary experiments demonstrate the applicability of \approach to larger qubit counts and more complex quantum circuits than state-of-the-art repair approaches. Initial results show a repair rate of 70\% across both real and artificially generated faults and provide valuable insights for manual repair in cases where a fault was not repaired. 
\looseness=-1

\bibliographystyle{IEEEtran}
\bibliography{references}

\end{document}